\documentclass{article}
\usepackage{epsfig}
\usepackage{amsmath}
\usepackage{amsfonts}
\usepackage{amssymb}
\usepackage{amscd}
\usepackage{graphics}
\usepackage{color}

\def\x{{\sf x}}

\def\a{{\sf a}}

\def\X{{\sf X}}

\newcommand{\beq}{\begin{equation}}
\newcommand{\eeq}{\end{equation}}
\newcommand{\beqa}{\begin{eqnarray}}
\newcommand{\eeqa}{\end{eqnarray}}
\begin{document}

\title{ Euclidean Relativistic Quantum Mechanics}
\author{W. N. Polyzou,\\
Philip Kopp\\
Department of Physics and Astronomy,\\ 
The University of Iowa, Iowa  City, IA 52242}
\maketitle
\begin{abstract} 
  We discuss a formulation of exactly Poincar\'e invariant quantum
  mechanics where the input is model Euclidean Green functions or
  their generating functional.  We discuss the structure of the
  models, the construction of the Hilbert space, the construction and
  transformation properties of single-particle states, and the
  construction of GeV scale transition matrix elements.  A simple
  model is utilized to demonstrate the feasibility of this approach.
\end{abstract}

\section{Introduction}

The motivation for this work is to construct mathematically
well-defined quantum mechanical models of few-body systems at the GeV
energy scale that have a direct relation to an underlying quantum
field theory. The goal is to use experience gained from the field
theory to constrain the structure of the models.

We do this by starting with the quantum mechanical interpretation of
the field theory.  Normally this is given in terms of vacuum expectation
values of products of fields (Wightman
functions), which represent the kernel of the Hilbert space inner
product of the field theory \cite{Wightman:1980}; however the Wightman
functions do not have a simple connection with the Lagrangian
formulation of the field theory.  The Euclidean Green functions have
the advantage that they can be directly related to Lagrangian
field theory and at the same time can be used to
reconstruct the underlying quantum theory
\cite{Osterwalder:1973dx}\cite{Osterwalder:1974tc}\cite{Frohlich:1974}.

With this connection in mind we consider a class of models that are
ideally expressed in terms of Euclidean-invariant reflection-positive
Green functions or their generating functional.  The generating functional
is formally the functional Fourier transform of the path measure:  
\begin{equation}
Z[f] := \int D_e[\phi] e^{-{A}[\phi] + i \phi (f) }   
= \sum_{n} {(i)^n \over n!} G_n \underbrace{(f,\cdots,f)}_{\mbox{n times}} 
=\prod_n \exp \left ( {i^n \over n!} C_n (f,\cdots ,f)\right) .
\label{eq.1}
\end{equation}
This provides the formal relation to the field-theoretic Lagrangian.
For the purpose of illustration we restrict our considerations to
generating functionals for scalar fields.  The $G_n$ are the $n$-point
Euclidean Green functions smeared over test functions in Euclidean
space-time variables and the $C_n$ are the corresponding connected
Green functions. 

The generating functionals are assumed to be Euclidean invariant, reflection
positive, and satisfy space-like cluster properties.  The test functions are
assumed to be Schwartz functions with support for positive Euclidean time.
We denote this space by ${\cal S}^+$.
Euclidean transforms on the test functions are denoted by 
$f(\x) \to f_{O,\a}(\x) := f(O \x +\a) )$ and Euclidean time reflection is 
denoted by $\Theta f(\tau , \mathbf{x})  :=  f(-\tau , \mathbf{x})$. 
The requirements on the generating functional are
\cite{Frohlich:1974}\cite{glimm:1981}:
\begin{equation}
Z[f] = Z[f_{O,\a}] \qquad \mbox{\bf Euclidean invariance}
\label{eq.2}
\end{equation}
\begin{equation}
\{ f_i \}_{i=1}^N \in {\cal S}^+ \qquad 
M_{ij} = Z[f_i - \Theta f_j ] \geq 0
\qquad \mbox{\bf reflection positivity}
\label{eq.3}
\end{equation}
\begin{equation}
\lim_{\vert \mathbf{a}\vert  \to \infty} \left ( Z[f+g_{I,\mathbf{a}}] - Z[f] Z[g] \right ) \to 0 
\qquad \mbox{\bf cluster properties}.
\label{eq.4}
\end{equation}
Models can be constructed by specifying a set of model connected
$n$-point functions, $C_n$ in (\ref{eq.1}).  A sufficient condition
for reflection positivity is that each term in the product
(\ref{eq.1}) is reflection positive.

A dense set of vectors in the model Hilbert space is given by
functionals of the form:
\begin{equation}
B[\phi]= \sum_{j=1}^{N_b} b_j e^{i \phi (f_j)} \qquad
C[\phi]= \sum_{k=1}^{N_c} c_k e^{i \phi (g_k)}
\label{eq.5}
\end{equation}
where
$b_j,c_k\in \mathbb{C}$,  $f_j,g_k \in {\cal S}_+$ and
$N_b,N_c < \infty$.
The model Hilbert inner product of two such vectors is 
\begin{equation}
\langle B \vert C \rangle :=
\sum_{j=1}^{N_b} \sum_{k=1}^{N_c} 
b^*_j c_k Z[ g_k- \Theta f_j ] =
\int D_e[\phi] e^{-A[\phi]} B^*[\phi \circ \theta ]C[\phi ] .
\label{eq.6}
\end{equation}
The representation at the end of eq. (\ref{eq.6}) suggests that we can
think of the vectors as wave functionals,  however the 
computation of the inner product only requires the generating functional.
The reflection positivity
condition ensures that vectors have positive length:
\begin{equation}
\langle B \vert B \rangle \geq 0 . 
\label{eq.7}
\end{equation}

To understand how Poincar\'e invariance is realized observe that the 
determinants of the following matrices are ($-$) the squares of the 
Lorentz and Euclidean lengths respectively:  
\begin{equation}
X=
\left ( 
\begin{array}{cc} 
t+z & x-iy \\
x+iy & t-z 
\end{array} 
\right ) 
\qquad
\X = \left ( 
\begin{array}{cc} 
i \tau+z & x-iy \\
x+iy & i \tau -z 
\end{array} 
\right ) .
\label{eq.8}
\end{equation}
The group of linear transformation that preserves both of these
determinants is $SL(2,\mathbb{C}) \times SL(2,\mathbb{C})$:
\begin{equation}
\X' = A \X B^t \qquad X = A X B^t  \qquad \det (A) = \det (B) =1 .
\label{eq.9}
\end{equation}
These represent complex Lorentz or complex orthogonal
transformations.  Real Lorentz transformations have $B=A^{*}$ while
real orthogonal transformations have $A,B \in SU(2)$.  The group of
real orthogonal transformations form a subgroup of the complex Lorentz
group in the inner product (\ref{eq.6}).  When one accounts for the
support condition on the test functions, Euclidean time evolution
becomes a contractive semigroup, rotations in Euclidean space-time
planes become local symmetric semigroups \cite{Klein:1981}\cite{Klein:1983}
\cite{Frohlich:1983kp}, and translations in a fixed direction and
rotations about a fixed axis become unitary one-parameter groups.  
The generators of
these transformations are self-adjoint operators on the physical
Hilbert space.  The one-parameter groups (semigroups) can be expressed
in terms of their infinitesimal generators as
\begin{equation}
e^{-\beta H} \qquad \beta > 0
\qquad
e^{ i \mathbf{a} \cdot \mathbf{P}}
\qquad 
e^{i \mathbf{J} \cdot \hat{\mathbf{n}} \psi }
\qquad
e^{ \mathbf{K} \cdot \hat{\mathbf{n}}\psi}.  
\label{eq.10}
\end{equation}
It is straightforward to show that the generators 
$\{ H,\mathbf{P}, \mathbf{J}, \mathbf{K} \}$ satisfy the
commutation relations of the Poincar\'e Lie algebra.

In this framework particles are point spectrum eigenstates of the 
square of the mass operator:
$M^2:= H^2 - \mathbf{P}^2$.   Normalizable mass eigenstates can be 
represented as wave functionals 
\begin{equation}
B_\lambda  [\phi] = 
\sum_n b_n  e^{i \phi (f_{n})} .
\end{equation}
Simultaneous eigenfunctionals of mass, linear momentum and angular
momentum can be extracted from these mass eigenstates using rotations
and translations:
\begin{equation}
B_\lambda (\mathbf{p}) [\phi] = 
\int {d^3a \over (2 \pi)^{3/2}}  e^{-i \mathbf{p}\cdot \mathbf{a}}  
B_{\lambda,I,\mathbf{a}} [\phi] 
\label{eq.11}
\end{equation}
\begin{equation}
B_{\lambda,j} (\mathbf{p},\mu ) [\phi] := 
\int_{SU(2)} dR  \sum_{\nu=-j}^j 
B_{\lambda,R,0} (R^{-1} \mathbf{p})
[\phi]D^{j*}_{\mu \nu} (R).  
\label{eq.12}
\end{equation}
The single-particle wave functionals can be interpreted as
multiplication operators.  These single-particle wave functionals can
be used to construct the two Hilbert space injection operators that
define the asymptotic conditions in the two Hilbert space
\cite{Coester:1965zz} formulation of Haag-Ruelle Scattering theory
\cite{Haag:1958vt}\cite{Ruelle:1962}
\cite{simon}\cite{baumgartl:1983}.  The wave operators and injection
operator have the form

\begin{equation}
\vert \Psi_{\pm} (f_1, \cdots f_n) \rangle : = 
\lim_{t \to \infty} e^{iHt} \Phi e^{-i H_0t} \vert \mathbf{f} \rangle =
\Omega_{\pm} \vert \mathbf{f} \rangle
\label{eq.13}
\end{equation}
\[
\Phi \vert \mathbf{f} \rangle [\phi]  = 
\]
\begin{equation}
\int  
\sum \prod_k \left ( \omega_{\lambda_k}(\mathbf{p}_k)
B_{\lambda_k,j_k} (\mathbf{p_k},\mu_k ) [\phi]
 -[H, B_{\lambda_k,j_k} (\mathbf{p}_k,\mu_k ) [\phi]]   
\right ) 
\tilde{f}_k(\mathbf{p}_k,\mu_k)   
d\mathbf{p}_k .
\label{eq.14}
\end{equation}
The asymptotic Hilbert space is the tensor product of 
one-particle irreducible representation spaces of the Poincar\'e group.  
Existence of the wave operators can be checked in a
given model by verifying the finiteness of the integral
\cite{cook}:
\begin{equation}
\int_0^{\pm \infty} \Vert (H\Phi-\Phi H_0) e^{-i H_0t} \vert \mathbf{f} 
\rangle \Vert dt 
< \infty ,
\end{equation}
while Poincar\'e covariance of the wave operators,
\begin{equation}
U(\Lambda ,a) \Omega_{\pm} = \Omega_{\pm} U_0 (\Lambda ,a)  
\label{eq.15}
\end{equation}
can be checked by verifying the asymptotic condition in this
representation of the Hilbert space
\begin{equation}
\lim_{t \to \pm \infty} \Vert (\mathbf{K}\Phi-\Phi \mathbf{K}_0)                
e^{-i H_0t} \vert \mathbf{f}                                                    
\rangle \Vert =0 .                                                                
\label{eq.16}
\end{equation}

Approximate sharp-momentum transition matrix elements can be computed 
from $S$ matrix elements in narrow wave packets using \cite{Brenig:1959}
\begin{equation}
\langle \mathbf{p}_1' ,\mu_1', \cdots , \mathbf{p}_n', \mu_n' \vert  
T \vert \mathbf{p}_1 ,\mu_1,\mathbf{p}_2, \mu_2 \rangle
\approx  
{\langle \mathbf{f}_{f} \vert S \vert \mathbf{f}_{i} \rangle
-
\delta_{ab}\langle \mathbf{f}_{f} \vert \mathbf{f}_{i} \rangle
\over 2 \pi i 
\langle \mathbf{f}_{f} \vert \delta (E_+-E_-) \vert \mathbf{f}_{i} \rangle }.
\label{eq.17}
\end{equation}
Using the Kato-Birman invariance principle \cite{kato:1966}
\cite{Chandler:1976}\cite{simon}\cite{baumgartl:1983}
the expression for the 
wave operators can be replaced by the limits
\begin{equation}
\Omega_{\pm}:=  \lim_{t \to \pm \infty}e^{-iHt}\Phi e^{iH_0t} =
\lim_{n \to \pm \infty} e^{in e^{-\beta H} } \Phi e^{-i n e^{-\beta H_0} }.
\label{eq.18}
\end{equation}
which for large enough $n$ gives the approximate expression for the 
$S$-matrix elements in normalizable states:
\begin{equation}
\langle \mathbf{f}_{f} \vert S \vert \mathbf{f}_{i} \rangle =
\langle \mathbf{f}_{f} \vert \Omega_+^{\dagger}\Omega_-   \vert \mathbf{f}_{i} \rangle
\approx \langle \mathbf{f}_{f} \vert e^{-in e^{-\beta H_f}} 
\Phi^{\dagger}  e^{2in e^{-\beta H}} \Phi e^{-in e^{-\beta H_f}} 
\vert \mathbf{f}_{i} \rangle .
\label{eq.19}
\end{equation}
The compactness of the spectrum of $\exp(-\beta H)$ means that  
for large but fixed $n$ that $e^{2in e^{-\beta H}}$ can be uniformly approximated 
by polynomial in $\exp (- \beta H)$:
\begin{equation}
e^{2in e^{-\beta H}} \approx \sum c_m(n) (e^{-\beta m H}) .
\label{eq.20}
\end{equation}
Chebyshev expansions provide an accurate approximation 
\cite{oxford2004}
for large
values of $n$:
\begin{equation}
f(e^{-\beta H}) \approx {1 \over 2} c_0 T_0 (e^{-\beta H}) + \sum_{k=1}^N c_k T_k (e^{-\beta H}) 
\label{eq.21}
\end{equation}
\begin{equation}
c_j = {2 \over N+1} \sum_{k=1}^N f( \cos({2k-1 \over N+1}{\pi
\over 2}) \cos(j {2k-1 \over N+1}{\pi \over 2}). 
\label{eq.22}
\end{equation}
We demonstrate the feasibiliy of this computational method using an
exactly solvable relativistic model with a mass square operator
given by 
\begin{equation}
M^2 = 4(\mathbf{k}^2+ m^2) - \vert g \rangle \lambda \langle g \vert 
\label{eq.23}
\end{equation}
\begin{equation}
\langle \mathbf{k} \vert g \rangle = {1 \over m_{\pi}^2 + \mathbf{k}^2}
\label{eq.24}
\end{equation}
where $m$ is mass of a nucleon and $\lambda$ is chosen to give a 
bound state with the mass of a deuteron.
First we test the approximation in equation (\ref{eq.17}). 
We use Gaussian
wave packets of the form $e^{-\alpha (k-k_0)^2}$ and find that to get 
sharp-momentum transition matrix
elements to a 0.1\% accuracy the width of the wave packet,
$k_w=1/\sqrt{\alpha}$, needs to be
about 3\% of the initial momentum, $k_0$.  This works at least 
up to 2 GeV.
The results are illustrated in table 1: 
\begin{center}
\vbox{
\begin{center}
Table 1
\end{center}
\begin{center}
\begin{tabular}{lllll}
\hline
$k_0$ & $\alpha$ &  $k_w$  &    \% error &  $k_w/k_0$ \\
\hline
[GeV] & [GeV$^{-2}]$ & [GeV] &   &  \\
\hline
0.1  &105000	&0.00308607	&0.1     &0.030 \\
0.3  &10500	&0.009759	&0.1     &0.032 \\
0.5  &3000	&0.0182574	&0.1     &0.036 \\
0.7  &1350	&0.0272166	&0.1     &0.038 \\
0.9  &750	&0.0365148	&0.1     &0.040 \\
1.1  &475	&0.0458831	&0.1     &0.041 \\
1.3  &330	&0.0550482	&0.1     &0.042 \\
1.5  &250	&0.0632456	&0.1     &0.042 \\
1.7  &190	&0.0725476	&0.1     &0.042 \\
1.9  &150	&0.0816497	&0.1     &0.042 \\
\hline
\end{tabular}
\end{center}
}
\end{center}
Next we test the approximation in (\ref{eq.19}) for the wave packet
widths in table 1.  We choose $\beta$ so $\beta$ times the center of 
momentum (CM) energy is a number of order unity. 
Table 2 shows that for $n=300$ we get ten figure
accuracy in the real and imaginary parts of the $S$-matrix elements
for a  2GeV incident CM momentum.  Similar results are obtained for 
all momentum scales between 100 MeV and 1.9 GeV.   
\vbox{
\begin{center}
{\bf Table 2: $k_0=2.0$[GeV], $\alpha=135$[GeV$^{-2}$]  }
\end{center}
\begin{center}
\begin{tabular}{lll}
\hline
$n$ & Re $\langle \phi \vert (S_n-I) \vert \phi \rangle$ & Im $\langle \phi \vert (S_n-I) \vert \phi \rangle$ \\
\hline
50  & -2.60094316473225e-6 & 1.94120750171791e-3 \\ 
100 & -2.82916859895010e-6 & 2.35553585404449e-3 \\
150 & -2.83171624670953e-6 & 2.37471383801820e-3 \\ 
200 & -2.83165946257657e-6 & 2.37492460997990e-3 \\ 
250 & -2.83165905312632e-6 & 2.37492527186858e-3 \\ 
300 & -2.83165905257121e-6 & 2.37492527262432e-3 \\ 
350 & -2.83165905190508e-6 & 2.37492527262493e-3 \\
400 & -2.83165905234917e-6 & 2.37492527262540e-3 \\
\hline                                             
ex  & -2.83165905227843e-6 & 2.37492527259701e-3 \\
\hline
\end{tabular}
\end{center}
}
Finally we test the Cheybshev approximation for the wave packet widths
in table 1 and the $n$-values in table 2.  Table 3 shows that for polynomials
of degree slightly larger than $n$ one obtains a 10-13 figure accuracy 
uniformly for spectrum of $\mbox{exp}(-\beta H)$.  
\begin{center}
\vbox{
\begin{center}
{\bf 
Table 3: Convergence with respect to Polynomial degree $e^{inx}$
}
\end{center}
\begin{center}
\begin{tabular}{llll}
\hline
x     &  n  &    deg   &    poly error \% \\
\hline
0.1  &   200  &  200 &   3.276e+00    \\
0.1  &   200  &  250 &   1.925e-11  \\
0.1  &   200  &  300 &   4.903e-13  \\
\hline
0.1  &   630  &  630 &   2.069e+00    \\
0.1  &   630  &  680 &   5.015e-08   \\
0.1  &   630  &  700 &   7.456e-11   \\
\hline                         
0.5  &   200  &  200 &   1.627e-13  \\
0.5  &   200  &  250 &   3.266e-13  \\
\hline0.5  &   630  &  580 &   1.430e-14  \\
0.5  &   630  &  680 &   9.330e-13  \\
\hline                
0.9  &   200  &  200 &   3.276e+00      \\
0.9  &   200  &  250 &   1.950e-11  \\
0.9  &   200  &  300 &   9.828e-13  \\
\hline
0.9  &   630  &  630 &   2.069e+00  \\
0.9  &   630  &	 680 &   5.015e-08  \\
0.9  &	 630  &  700 &   7.230e-11  \\
\hline
\end{tabular}
\end{center}
}
\end{center}
Table 4 shows the final approximation for the real and imaginary 
parts of the sharp-momentum transition matrix elements for CM
momenta up to 1.9 GeV.  The results are all within less than 
$0.1\%$ of the exact results in this model.
\begin{center}
\vbox{
\begin{center}
{\bf 
Table 4: Approximate transition matrix elements
}
\end{center}
\begin{center}
\begin{tabular}{llll}
\hline
$k_0$ & Real T & Im T & \% error \\
\hline
0.1   &  -2.30337e-1    &  -4.09325e-1	          &     0.0956   \\
0.3   &  -3.46973e-2    &  -6.97209e-3	          &     0.0966   \\
0.5   &  -6.44255e-3    &  -3.86459e-4	          &     0.0986   \\
0.7   &  -1.88847e-3    &  -4.63489e-5           &      0.0977   \\
0.9   &  -7.28609e-4    &  -8.86653e-6           &      0.0982   \\
1.1   &  -3.35731e-4    &  -2.30067e-6           &      0.0987   \\
1.3   &  -1.74947e-4    &  -7.38285e-7           &      0.0985   \\
1.5   &  -9.97346e-5    &  -2.76849e-7           &      0.0956   \\
1.7   &  -6.08794e-5    &  -1.16909e-7           &      0.0964   \\
1.9   &  -3.92110e-5    &  -5.42037e-8           &      0.0967   \\
\hline                                                                          \end{tabular}
\end{center}
}
\end{center}
\section{Conclusion}

We presented a formulation of relativistic quantum mechanics
\cite{Wigner:1939cj} that uses Euclidean generating functionals or
Green functions as input.  In applications these have to be
modeled. One virtue of this representation is that all calculations
can be performed without analytic continuation.

The expression in equation (\ref{eq.1}) suggests that the generating
functionals can be modeled using a finite collection of model
connected Green functions.  While it is easy to maintain Euclidean
covariance and cluster properties of the models in this
representation, reflection positivity is a non-trivial constraint that
will be the subject of future investigations.  While it holds for free
field generating functionals, it is not stable with respect to small
perturbations \cite{Wessels:2003af}.  Failure of reflection positivity
points to violations of the spectral condition or the positivity of
the Hilbert space norms.

The model calculations presented suggest that for models based on
reflection positive generating functionals this framework can be used
to accurately compute both bound state and scattering observables.

This work supported in part by the U.S. Department of Energy, under 
contract DE-FG02-86ER40286.

\end{document}